\documentclass[10pt,twocolumn,english,manuscript]{revtex4}
\usepackage[T1]{fontenc}
\usepackage[latin9]{inputenc}
\usepackage{amsmath}
\usepackage{amssymb}
\usepackage{graphicx}

\makeatletter
\@ifundefined{textcolor}{}
{%
 \definecolor{BLACK}{gray}{0}
 \definecolor{WHITE}{gray}{1}
 \definecolor{RED}{rgb}{1,0,0}
 \definecolor{GREEN}{rgb}{0,1,0}
 \definecolor{BLUE}{rgb}{0,0,1}
 \definecolor{CYAN}{cmyk}{1,0,0,0}
 \definecolor{MAGENTA}{cmyk}{0,1,0,0}
 \definecolor{YELLOW}{cmyk}{0,0,1,0}
}

\usepackage{babel}

\makeatother

\usepackage{babel}
\begin{document}

\title{Thermal entanglement in a spin-1/2 Ising-XYZ distorted diamond chain with the second-neighbor interaction between nodal Ising spins}

\author{Onofre Rojas$^{1}$, M. Rojas$^{1}$, S. M. de Souza$^{1}$, J. Torrico$^{2}$, J. Stre\v{c}ka$^{3}$ and M. L. Lyra$^{2}$}

\affiliation{$^{1}$Departamento de Física, Universidade Federal de Lavras, 37200-000, Lavras-MG, Brazil}

\affiliation{$^{2}$Instituto de Física, Universidade Federal de Alagoas, 57072-970, Maceio, AL, Brazil}

\affiliation{$^{3}$Department of Theoretical Physics and Astrophysics, Faculty of Science, P.J. \v{S}af\'arik University,
Park Angelinum 9, 040 01 Ko\v{s}ice, Slovak Republic}
\begin{abstract}
We consider a spin-1/2 Ising-XYZ distorted diamond chain with the XYZ interaction between the interstitial Heisenberg dimers, the nearest-neighbor Ising coupling between the nodal and interstitial spins, respectively, and the second-neighbor Ising coupling between the nodal spins. The ground-state phase diagram of the spin-1/2 Ising-XYZ distorted diamond chain exhibits several intriguing phases due to the XY anisotropy and the second-neighbor interaction, whereas the model can be exactly solved using the transfer-matrix technique. The quantum entanglement within the Heisenberg spin dimers is studied through a bipartite measure concurrence, which is calculated from a relevant reduced density operator. The concurrence may either show a standard thermal dependence with a monotonous decline with increasing temperature or a more peculiar thermal dependence accompanied with reentrant behavior of the concurrence. It is conjectured that the bipartite entanglement between the interstitial Heisenberg spin pairs in the natural mineral azurite is quite insensitive to the applied magnetic field and it persists up to approximately 30 Kelvins.
\end{abstract}
\maketitle

\section{Introduction}

Over the last few decades, quantum properties of low-dimensional spin systems are one of topical issues of modern condensed matter physics \cite{AmicoHorod}. In this research field many attempts have been dedicated first of all to a qualitative and quantitative characterization of quantum entanglement in solid-state magnetic materials, which provide a potential physical resource for quantum computation, quantum cryptography and quantum communication \cite{nielsen}. From this perspective, the quantum entanglement has been particularly studied within the Heisenberg spin chains \cite{qubit-Heisnb}, because they belong to the simplest exactly tractable quantum spin systems with a non-trivial entanglement between constituent spins.

Curiously, magnetic properties of several insulating magnetic materials can be satisfactorily described by one-dimensional quantum Heisenberg spin models. For instance, it is quite interesting to consider the spin-1/2 quantum Heisenberg model on a distorted diamond chain, since this model captures properties of real magnetic materials such as $\mathrm{Cu_{3}(CO_{3})_{2}(OH)_{2}}$ known as the natural mineral azurite \cite{kikuchi05}. The outcomes of experimental measurements on the azurite were testified by solving the relevant quantum Heisenberg model through several complementary approaches \cite{Lau}. It is noteworthy that Honecker \textit{et al}. \cite{honecker} has brought a deeper insight into magnetic properties of the azurite when calculating thermodynamic and dynamic properties of the spin-1/2 Heisenberg model on a distorted diamond chain by adapting of powerful numerical methods. On the other hand, magnetic properties and thermodynamics of the simplified Ising-Heisenberg diamond chains were widely discussed on the grounds of rigorous analytical calculations \cite{Cano,valverde,orojas,lis-1}.

Recently, a considerable attention has been therefore paid to the Ising-Heisenberg diamond chains. Although
these theoretical models are subject to a substantial simplification of the physical reality,
the exactly solvable Ising-Heisenberg spin chains surprisingly provide a reasonable quantitative
description of the magnetic behavior associated with the real spin-chain
materials. The thermal entanglement in some exactly solvable Ising-Heisenberg
diamond chains was investigated in Refs. \cite{mrojas,mrojas-1}. Later, the fermionic entanglement
measures were calculated for the hybrid diamond chain composed of the localized Ising spins and mobile electrons \cite{mrojas2}.
Inspired by these works, the quantum teleportation of two qubits for an arbitrary pure
entangled state via two independent Ising-XXZ diamond chains considered as a quantum channel
has been investigated in Ref. \cite{moises-aphy}. The quantities such as the output entanglement, fidelity and average fidelity of teleportation
were studied by assuming the system in thermal equilibrium \cite{moises-aphy}. Very recently, the finite-temperature
scaling of trace distance discord near criticality of the Ising-XXZ diamond chain was also examined in detail \cite{cheng}.

The generalized spin-1/2 Ising-Heisenberg diamond chain has been discussed
by Lisnyi \textit{et al}. in Ref. \cite{stre}, where they demonstrated that
the second-neighbor interaction between the nodal spins is responsible
for emergence of intermediate plateaux in low-temperature magnetization curves.
A similar model also has been considered by Faizi and Eftekhari \cite{Faizi}, where the concurrence,
1-norm geometric quantum discord and quantum discord of the diamond chain were compared.
Later, the spin-1 Ising-Heisenberg diamond chain with the second-neighbor interaction
between nodal spins has been discussed in Ref. \cite{vahan}, which evidences existence
of three novel quantum ground states with a translationally broken symmetry though the total number
of plateaux in a zero-temperature magnetization remains the same.

The paper is organized as follows. In Sec. II we present the spin-1/2 Ising-XYZ
distorted diamond chain with the second-neighbor interaction between the nodal spins along with
a brief discussion of the zero-temperature phase diagram. In Sec. III, we obtain
the exact solution of the model via the transfer-matrix approach, which allows
a straightforward calculation of the reduced density operator. In Sec. IV, we discuss
the thermal entanglement of the investigated model such as the concurrence and the threshold temperature.
Finally, our conclusions are presented in Sec. V.

\section{Ising-XYZ diamond chain }

\begin{figure}
\includegraphics[scale=0.5]{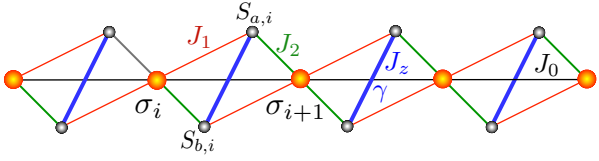}\caption{\label{fig:diamond}(Color online) A schematic representation of the asymmetric
spin-1/2 Ising-XYZ diamond chain. The Ising spins are denoted by $\sigma_{i}$ and the Heisenberg spins are represented by $S_{a(b),i}$.  }
\end{figure}

In this work, we will consider the spin-1/2 Ising-XYZ distorted diamond chain with the second-neighbor
interaction between nodal spins as schematically depicted in Fig.\ref{fig:diamond}. The investigated spin
system is composed of the Ising spins $\sigma_{i}$ located at the nodal lattice sites and
the Heisenberg spin pairs $S_{a,i}$ and $S_{b,i}$ located at each couple of the interstitial sites (see Fig.\ref{fig:diamond}).
The total Hamiltonian of the spin-1/2 Ising-XYZ distorted diamond chain can be written as a sum of block Hamiltonians
\begin{align}
\mathcal{H}= & \sum_{i=1}^{N}\mathcal{H}_{i},
\end{align}
whereas the $i$-th block Hamiltonian can be defined as
\begin{align}
\mathcal{H}_{i}= & J(1+\gamma)S_{a,i}^{x}S_{b,i}^{x}+J(1-\gamma)S_{a,i}^{y}S_{b,i}^{y}+J_{z}S_{a,i}^{z}S_{b,i}^{z}\nonumber \\
 & +\sigma_{i}\left(J_{1}S_{a,i}^{z}+J_{2}S_{b,i}^{z}\right)+\sigma_{i+1}\left(J_{2}S_{a,i}^{z}+J_{1}S_{b,i}^{z}\right)\nonumber \\
 & +J_{0}\sigma_{i}\sigma_{i+1}-h\left(S_{a,i}^{z}+S_{b,i}^{z}\right)-\frac{h}{2}\left(\sigma_{i}+\sigma_{i+1}\right).
 \label{eq:Ham-1}
\end{align}
Here, the coupling constants $J$, $\gamma$ and $J_{z}$ denote the $XYZ$ interaction within the interstitial Heisenberg dimers,
the coupling constants $J_{1}$ and $J_{2}$ correspond to the nearest-neighbor Ising interaction between the nodal and interstitial spins,
while the coupling constant $J_{0}$ represents the second-neighbor Ising interaction between the nodal spins. Finally, the parameter
$h = g \mu_B B$ incorporates the effect of external magnetic field $B$ applied along the $z$-axis, $g$ is the respective gyromagnetic ratio
and $\mu_B$ is Bohr magneton.

After a straightforward calculation one can obtain the eigenvalues of the block Hamiltonian \eqref{eq:Ham-1}
\begin{align}
\varepsilon_{i1,i4}= & \frac{J_{z}}{4}+J_{0}\sigma_{i}\sigma_{i+1}-\frac{h}{2}(\sigma_{i}+\sigma_{i+1})\pm\frac{\nu_{\sigma_{i}\sigma_{i+1}}}{2},\label{eq:Eg14}\\
\varepsilon_{i2,i3}= & -\frac{J_{z}}{4}+J_{0}\sigma_{i}\sigma_{i+1}-\frac{h}{2}(\sigma_{i}+\sigma_{i+1})\pm\frac{\bar{\nu}_{\sigma_{i}\sigma_{i+1}}}{2},\label{eq:Eg23}
\end{align}
with
\begin{alignat}{1}
\nu_{\sigma_{i}\sigma_{i+1}} & =\sqrt{J^{2}\gamma^{2}+[(J_{1}+J_{2})(\sigma_{i}+\sigma_{i+1})-2h]^{2}},\label{eq:v}\\
\bar{\nu}_{\sigma_{i}\sigma_{i+1}} & =\sqrt{J^{2}+(J_{1}-J_{2})^{2}(\sigma_{i}-\sigma_{i+1})^{2}}.\label{eq:vb}
\end{alignat}
The corresponding eigenvectors of the cell Hamiltonian \eqref{eq:Ham-1} in the standard
dimer basis $\{|\uparrow\uparrow\rangle_i,|\uparrow\downarrow\rangle_i,|\downarrow\uparrow\rangle_i,|\downarrow\downarrow\rangle_i\}$
are given respectively by
\begin{alignat}{1}
|\tau_{1}\rangle_i= & \cos(\tfrac{\phi_i}{2})|\uparrow\uparrow\rangle_i+\sin(\tfrac{\phi_i}{2})|\downarrow\downarrow\rangle_i,\\
|\tau_{2}\rangle_i= & \cos(\tfrac{\varphi_i}{2})|\uparrow\downarrow\rangle_i+\sin(\tfrac{\varphi_i}{2})|\downarrow\uparrow\rangle_i,\\
|\tau_{3}\rangle_i= & \sin(\tfrac{\varphi_i}{2})|\uparrow\downarrow\rangle_i-\cos(\tfrac{\varphi_i}{2})|\downarrow\uparrow\rangle_i,\\
|\tau_{4}\rangle_i= & \sin(\tfrac{\phi_i}{2})|\uparrow\uparrow\rangle_i-\cos(\tfrac{\phi_i}{2})|\downarrow\downarrow\rangle_i,
\end{alignat}
where the probability amplitudes $\phi_i$ and $\varphi_i$ are expressed through the mixing angle
\begin{equation}
\phi_i=\tan^{-1}\left(\tfrac{J\gamma}{(J_{1}+J_{2})(\sigma_{i}+\sigma_{i+1})-2h}\right),\label{eq:phi}
\end{equation}
with $0\leqslant\phi_i\leqslant2\pi$ and
\begin{equation}
\varphi_i=\tan^{-1}\left(\tfrac{J}{(J_{1}-J_{2})(\sigma_{i}-\sigma_{i+1})}\right),\label{eq:vphi}
\end{equation}
with $0\leqslant\varphi_i\leqslant2\pi$.

Obviously, the mixing angles $\phi_i$ and $\varphi_i$ depend according to Eqs. (\ref{eq:phi}) and (\ref{eq:vphi}) on the nodal Ising spins $\sigma_{i}$ and $\sigma_{i+1}$ quite similarly as the corresponding eigenenergies $\varepsilon_{ij}$ given by Eqs. \eqref{eq:Eg14} and \eqref{eq:Eg23} do.

\subsection{Zero-temperature phase diagram}

\begin{figure}
\includegraphics[scale=0.32]{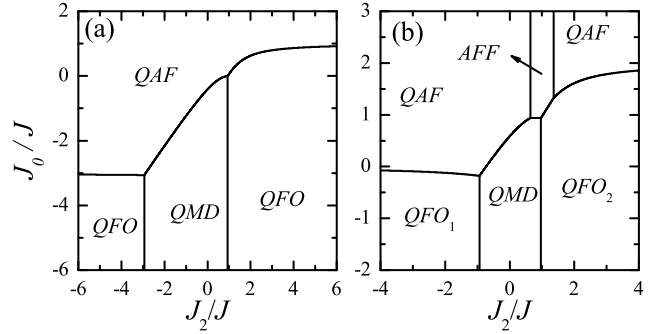}\vspace{-0.3cm}\caption{\label{fig:J0J2}The ground-state phase diagrams in the plane
$J_{2}/J$-$J_{0}/J$ by assuming the fixed values of the coupling constants $J_{1}/J=1$, $J_{z}/J=1$, $\gamma=0.5$ and two different magnetic fields:
(a) $h/J=0$; (b) $h/J=1$.}
\end{figure}

Now, let us discuss the ground-state phase diagram of the spin-1/2 Ising-$XYZ$
diamond chain with the second-neighbor interaction between the nodal spins.
In Fig.\ref{fig:J0J2}, we display the zero-temperature phase diagram in the plane
$J_{2}/J$-$J_{0}/J$ by assuming the fixed values of the coupling constants
$J_{1}/J=1$, $J_{_{z}}/J=1$ and $\gamma=0.5$. At zero magnetic field one detects
just three different ground states (Fig.\ref{fig:J0J2}a)
\begin{itemize}
\item The quantum antiferromagnetic phase $QAF$ with the antiferromagnetic alignment of the nodal Ising spins
and the quantum entanglement of two antiferromagnetic states of the Heisenberg dimers
\begin{equation}
\mid QAF\rangle=\overset{N}{\underset{i=1}{\prod}}\mid(-)^{i}\rangle_{i}\otimes\mid\tau_{3}\rangle_{i},
\end{equation}
whereas the corresponding ground-state energy per block is given by
\begin{alignat}{1}
E_{QAF}= & -\frac{J_{z}+J_{0}}{4}-\frac{1}{2}\sqrt{(J_{1}-J_{2})^{2}+J^{2}}.
\end{alignat}
\item The quantum ferromagnetic phases $QFO_1$ and $QFO_2$ with the classical ferromagnetic alignment of the nodal Ising spins and
the quantum entanglement of two ferromagnetic states of the interstitial Heisenberg dimers
\begin{alignat}{1}
\mid QFO_{1}\rangle= & \overset{N}{\underset{i=1}{\prod}}\mid+\rangle_{i}\otimes\mid\tau_{4}\rangle_{i},\label{eq:fmf1}\\
\mid QFO_{2}\rangle= & \overset{N}{\underset{i=1}{\prod}}\mid-\rangle_{i}\otimes\mid\tau_{4}\rangle_{i}.\label{eq:fmf2}
\end{alignat}
whereas the corresponding ground-state energy per block are given by
\begin{alignat}{1}
E_{QFO_{1}}= & \frac{J_{z}+J_{0}}{4}+\frac{h-\sqrt{(J_{1}+J_{2}+2h)^{2}+J^{2}\gamma^{2}}}{2},\\
E_{QFO_{2}}= & \frac{J_{z}+J_{0}}{4}-\frac{h+\sqrt{(J_{1}+J_{2}-2h)^{2}+J^{2}\gamma^{2}}}{2}.
\end{alignat}
Note that the eigenstates $QFO_{1}$ and $QFO_{2}$ have the same energy at zero magnetic field and hence,
this two-fold degenerate ground-state manifold is simply referred to as the $QFO$ state in the zero-field limit.
\item The quantum monomer-dimer phase ($QMD$) with the classical ferromagnetic alignment
of the nodal Ising spins and the fully entangled singlet state of the interstitial Heisenberg dimers
\begin{alignat}{1}
\mid QMD\rangle= & \overset{N}{\underset{i=1}{\prod}}\mid+\rangle_{i}\otimes\frac{1}{\sqrt{2}}(-\mid\uparrow\downarrow\rangle+\mid\downarrow\uparrow\rangle)_{i}
\end{alignat}
\end{itemize}
whereas its ground-state energy per block is
\begin{alignat}{1}
\mathcal{E}_{QMD}= & \frac{J_{0}-J_{z}}{4}-\frac{h}{2}-\frac{|J|}{2}.
\end{alignat}
The $QMD$ phase is wedged in the $QFO$ phase within the interval $-1-\sqrt{15}/2\leqslant J_{2}/J\leqslant-1+\sqrt{15}/2$,
while the phase boundary between the $QMD$ and $QAF$ phases is given by $J_{0}/J=1-\sqrt{(J_{2}/J-1)^{2}+1}$. Similarly,
the phase boundary between the $QAF$ and $QFO$ phases is given by $J_{0}/J=\sqrt{(J_{2}/J+1)^{2}+1/4}-\sqrt{(J_{2}/J-1)^{2}+1}-1$.

On the other hand, the ground-state phase diagram in the plane $J_{2}/J$-$J_{0}/J$ shown in Fig.\ref{fig:J0J2}b for the fixed value of the magnetic field $h/J=1$ additionally exhibits one more extra phase, which has character of the modulated antiferromagnetic-ferromagnetic phase $AFF$. More specifically, the $AFF$ ground state refers to the classical antiferromagnetic alignment of the nodal Ising spins accompanied with the quantum entanglement of two ferromagnetic states of the interstitial Heisenberg dimers
\begin{equation}
\mid AFF\rangle=\overset{N}{\underset{i=1}{\prod}}\mid(-)^{i}\rangle_{i}\otimes\mid\tau_{4}\rangle_{i},
\end{equation}
whereas the relevant ground-state energy per block reads
\[
E_{AFF}=-\frac{J_{0}}{4}+\frac{J_{z}}{4}-\frac{\sqrt{4h^{2}+J^{2}\gamma^{2}}}{2}.
\]
The $AFF$ phase is wedged in the $QAF$ ground state within the interval $1-\frac{1}{2}\sqrt{17-4\sqrt{17}}\leqslant J_{2}/J\leqslant1+\frac{1}{2}\sqrt{17-4\sqrt{17}}$, whereas the $QMD$ phase is wedged in between the $QFO_{1}$ and $QFO_{2}$ ground states within the interval $1-\sqrt{15}/2\leqslant J_2/J\leqslant-3+3\sqrt{7}/2$. It is worthwhile to recall that the two-fold degenerate ground-state manifold $QFO$ splits into two separate phases $QFO_{1}$ and $QFO_{2}$ given by the eigenvectors \eqref{eq:fmf1} and \eqref{eq:fmf2}, respectively, when the external magnetic field is turned on.

It has been demonstrated previously that the experimental data reported on the natural mineral azurite Cu$_{3}$(CO$_{3}$)$_{2}$(OH)$_{2}$ \cite{Lau,honecker} can be satisfactorily described by the spin-1/2 Heisenberg distorted diamond chain when assuming the following values of the coupling constants $J/k_B=J_{z}/k_B=33$~K, $J_{1}/k_B=15.5$~K, $J_{2}/k_B=6.9$~K, $J_{0}/k_B=4.6$~K, $\gamma=0$ and $g=2.06$.
\begin{figure}
\includegraphics[scale=0.22]{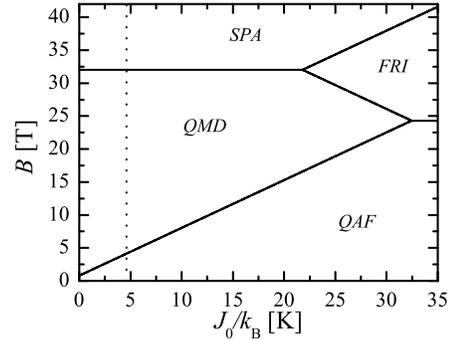}\vspace{-0.3cm}\caption{\label{fig:hJ0}The zero-temperature phase diagram in the $J_{0}/k_B$[K]-$B$[T] plane for the particular set of the coupling constants $J/k_B=J_{z}/k_B=33$~K, $J_{1}/k_B=15.5$~K, $J_{2}/k_B=6.9$~K, $\gamma=0$ and $g=2.06$. The vertical dotted line drawn for $J_{0}/k_B=4.6$K corresponds to a set of the coupling constants relevant to the azurite Cu$_{3}$(CO$_{3}$)$_{2}$(OH)$_{2}$.}
\end{figure}
To gain an insight into the magnetic behavior of the azurite Cu$_{3}$(CO$_{3}$)$_{2}$(OH)$_{2}$, we have therefore plotted in Fig.\ref{fig:hJ0} the ground-state phase diagram of the simplified spin-1/2 Ising-Heisenberg diamond chain in the $J_{0}/k_B-B$ plane. The dotted line shown for the particular value $J_{0}/k_B=4.6$~K of the second-neighbor interaction between the nodal spins corresponds to the same set of the coupling constants as being reported for the azurite \cite{Lau,honecker}. According to this plot, we have obtained two particular cases of the formerly described quantum ground states, namely, the saturated paramagnetic phase $SPA$ with a perfect ferromagnetic alignment of all Ising as well as Heisenberg spins
\begin{equation}
\mid SPA\rangle=\overset{N}{\underset{i=1}{\prod}}\mid+\rangle_{i}\otimes\mid\uparrow\uparrow\rangle_{i},
\end{equation}
which is retrieved from the $QFO_{1}$ phase (\ref{eq:fmf1}) in the particular limit $\phi=\pi$. Similarly, the classical ferrimagnetic phase $FRI$ with the antiferromagnetic alignment of the nodal Ising spins and the classical ferromagnetic alignment of the interstitial Heisenberg spins is retrieved from the modulated $AFF$ phase in the special limiting case $\phi=\pi$
\begin{equation}
\mid FRI\rangle=\overset{N}{\underset{i=1}{\prod}}\mid(-)^{i}\rangle_{i}\otimes\mid\uparrow\uparrow\rangle_{i}.
\end{equation}
The ground-state phase boundaries between the $QMD$-$SPA$ and $QAF$-$FRI$ phases appear at constant magnetic fields $B$[T] $\approx 31.9$ and $B$[T] $\approx 24.3$, respectively, while the remaining ground-state phase boundaries in Fig.\ref{fig:hJ0} are straight linear lines. The ground-state boundary between the $QAF$ and $QMD$ phases is given by $B$[T] $\approx 0.8 + 0.72J_{0}/k_B$, the ground-state boundary between the $QMD$ and $FRI$ phases is described by $B$[T] $\approx 47.7 - 0.72J_{0}/k_B$, while the ground-state boundary between the FRI and SPA phases follows from $B$[T] $\approx 22.4 + 0.72J_{0}/k_B$.

\section{Partition function and density operator}

In order to study thermal and magnetic properties of the spin-1/2 Ising-XYZ diamond chain we first need to obtain the partition function. As mentioned earlier \cite{Fisher,Syozi,phys-A-09}, the partition function of the spin-1/2 Ising-XYZ diamond chain can be exactly calculated through a decoration-iteration transformation and transfer-matrix approach. However, here we will summarize crucial steps of an alternative approach that allows a straightforward calculation of the reduced density operator
\begin{equation}
\varrho_i (\sigma_i,\sigma_{i+1})=\mathrm{e}^{-\beta\mathcal{H}_{i}(\sigma_i,\sigma_{i+1})},
\label{eq:rho-loc}
\end{equation}
where $\beta=1/(k_{B}T)$, $k_{B}$ is the Boltzmann's constant, $T$ is the absolute temperature and $\mathcal{H}_{i}(\sigma_i,\sigma_{i+1})$
corresponds to the $i$th-block Hamiltonian (\ref{eq:Ham-1}) depending on two nodal Ising spins $\sigma_i$ and $\sigma_{i+1}$.

Alternatively, the density operator (\ref{eq:rho-loc}) can be written in terms of the eigenvalues \eqref{eq:Eg14} and eigenvectors \eqref{eq:Eg23} of the cell Hamiltonian
\begin{equation}
\varrho_i (\sigma_i,\sigma_{i+1})=\sum_{j=1}^{4}\mathrm{e}^{-\beta\varepsilon_{ij}(\sigma_i,\sigma_{i+1})}|\varphi_{ij}\rangle\langle\varphi_{ij}|.
\end{equation}
By tracing out degrees of freedom of the $i$th Heisenberg dimer one straightforwardly obtains the Boltzmann factor
\begin{equation}
w (\sigma_i,\sigma_{i+1})=\mathrm{tr}_{ab}\left[\varrho_i (\sigma_i,\sigma_{i+1})\right]=\sum_{j=1}^{4}\mathrm{e}^{-\beta\varepsilon_{ij}(\sigma_i,\sigma_{i+1})}.
\label{eq:w-def}
\end{equation}
The partition function of the spin-1/2 Ising-XYZ diamond chain can be subsequently rewritten in terms of the associated Boltzmann's weights
\begin{equation}
Z_{N}=\sum_{\{\sigma\}}w(\sigma_{1},\sigma_{2})\ldots w(\sigma_{N},\sigma_{1}).
\end{equation}

Using the transfer-matrix approach, we can express the partition function of the spin-1/2 Ising-XYZ diamond chain as $Z_{N}=\mathrm{tr}\left(W^{N}\right)$, where the transfer matrix is defined as
\begin{equation}
W=\left[\begin{array}{cc}
w(\frac{1}{2},\frac{1}{2}) & w(\frac{1}{2},-\frac{1}{2})\\
w(-\frac{1}{2},\frac{1}{2}) & w(-\frac{1}{2},-\frac{1}{2})
\end{array}\right].\label{eq:W}
\end{equation}
For further convenience, the transfer-matrix elements are denoted as $w_{++}\equiv w(\frac{1}{2},\frac{1}{2})$,
$w_{+-}\equiv w(\frac{1}{2},-\frac{1}{2})$ and $w_{--}\equiv w(-\frac{1}{2},-\frac{1}{2})$ and they are explicitly
given by
\begin{alignat}{1}
w_{++}= & 2{\rm e}^{-\beta\left(\frac{J_{0}}{4}-\frac{h}{2}\right)}\left[{\rm e}^{-\frac{\beta J_{z}}{4}}{\rm ch}\left(\tfrac{\beta\nu_{++}}{2}\right)+{\rm e}^{\frac{\beta J_{z}}{4}}{\rm ch}\left(\tfrac{\beta|J|}{2}\right)\right],\nonumber \\
w_{--}= & 2{\rm e}^{-\beta\left(\frac{J_{0}}{4}+\frac{h}{2}\right)}\left[{\rm e}^{-\frac{\beta J_{z}}{4}}{\rm ch}\left(\tfrac{\beta\nu_{--}}{2}\right)+{\rm e}^{\frac{\beta J_{z}}{4}}{\rm ch}\left(\tfrac{\beta|J|}{2}\right)\right],\nonumber \\
w_{+-}= & 2{\rm e}^{\frac{\beta J_{0}}{4}}\left[{\rm e}^{-\frac{\beta J_{z}}{4}}{\rm ch}\left(\tfrac{\beta\nu_{+-}}{2}\right)+{\rm e}^{\frac{\beta J_{z}}{4}}{\rm ch}\left(\tfrac{\beta\bar{\nu}_{+-}}{2}\right)\right],\label{eq:ws}
\end{alignat}
where $\nu_{\sigma_i\sigma_{i+1}}$ and $\bar{\nu}_{\sigma_i\sigma_{i+1}}$ follow from Eqs.~\eqref{eq:v} and \eqref{eq:vb}, respectively.

After performing the diagonalization of the transfer matrix (\ref{eq:W}) one gets two eigenvalues
\begin{equation}
\Lambda_{\pm}=\frac{1}{2} \left[w_{++}+w_{--}\pm\sqrt{\left(w_{++}-w_{--}\right)^{2}+4w_{+-}^{2}} \right]
\label{eq:eig}
\end{equation}
and hence, the partition function of the spin-1/2 Ising-XYZ distorted diamond chain under periodic boundary conditions is given by
\begin{equation}
Z_{N}=\Lambda_{+}^{N}+\Lambda_{-}^{N}.
\end{equation}
In the thermodynamic limit $N \to \infty$, the free energy per unit cell is solely determined by the largest transfer-matrix eigenvalue
\begin{equation}
f=-\frac{1}{N\beta}\ln Z_{N}=-\frac{1}{\beta}\ln\Lambda_{+},\label{eq:free}
\end{equation}
which is explicitly given by Eqs. \eqref{eq:ws} and \eqref{eq:eig}.

\subsection{Reduced density operator in a matrix form}

To find the matrix representation of the reduced density operator depending on two nodal Ising spins $\sigma_i$ and $\sigma_{i+1}$ one may follow the procedure elaborated in Ref. \cite{moises-pra}. Thus, the matrix representation of the reduced density operator in the natural basis of the Heisenberg dimer reads
\begin{equation}
\varrho_i (\sigma_i,\sigma_{i+1})=\left(\begin{array}{cccc}
\varrho_{11} & 0 & 0 & \varrho_{14}\\
0 & \varrho_{22} & \varrho_{23} & 0\\
0 & \varrho_{32} & \varrho_{33} & 0\\
\varrho_{41} & 0 & 0 & \varrho_{44}
\end{array}\right),
\end{equation}
where the individual matrix elements are given by
\begin{align}
\varrho_{11} (\sigma_i,\sigma_{i+1}) = & \mathrm{e}^{-\beta\varepsilon_{i1}}\cos^{2}(\tfrac{\phi_i}{2})+\mathrm{e}^{-\beta\varepsilon_{i4}}\sin^{2}(\tfrac{\phi_i}{2}),\nonumber \\
\varrho_{14} (\sigma_i,\sigma_{i+1}) = & \left(\mathrm{e}^{-\beta\varepsilon_{i1}}-\mathrm{e}^{-\beta\varepsilon_{i4}}\right)\frac{\sin(\phi_i)}{2},\nonumber \\
\varrho_{22} (\sigma_i,\sigma_{i+1}) = & \mathrm{e}^{-\beta\varepsilon_{i2}}\cos^{2}(\tfrac{\varphi_i}{2})+\mathrm{e}^{-\beta\varepsilon_{i3}}\sin^{2}(\tfrac{\varphi_i}{2}),\nonumber \\
\varrho_{23} (\sigma_i,\sigma_{i+1}) = & \left(\mathrm{e}^{-\beta\varepsilon_{i2}}-\mathrm{e}^{-\beta\varepsilon_{i3}}\right)\frac{\sin(\varphi_i)}{2},\nonumber \\
\varrho_{33} (\sigma_i,\sigma_{i+1}) = & \mathrm{e}^{-\beta\varepsilon_{i2}}\sin^{2}(\tfrac{\varphi_i}{2})+\mathrm{e}^{-\beta\varepsilon_{i3}}\cos^{2}(\tfrac{\varphi_i}{2}),\nonumber \\
\varrho_{44} (\sigma_i,\sigma_{i+1}) = & \mathrm{e}^{-\beta\varepsilon_{i1}}\sin^{2}(\tfrac{\phi_i}{2})+\mathrm{e}^{-\beta\varepsilon_{i4}}\cos^{2}(\tfrac{\phi_i}{2}).
\end{align}
Notice that the eigenvalues $\varepsilon_{ij}$ are given by Eqs. \eqref{eq:Eg14} and \eqref{eq:Eg23}, whereas the mixing angles $\phi_i$ and $\varphi_i$ depending on the nodal Ising spins $\sigma_i$ and $\sigma_{i+1}$ are determined by Eqs. \eqref{eq:phi} and \eqref{eq:vphi}, respectively.

Next, the thermal average can be carried out for all except one Heisenberg dimer in order to construct the averaged reduced density operator \cite{moises-pra}. For this purpose, we will trace
out degrees of freedom of all interstitial Heisenberg dimers and nodal Ising spins except those from the $i$th block (unit cell) of a diamond chain. The transfer-matrix approach implies the following explicit expression for the partially averaged reduced density operator
\begin{align}
\rho_{i}= & \frac{1}{Z_{N}}\sum_{\{\sigma\}}w(\sigma_{1},\sigma_{2})\ldots w(\sigma_{i-1},\sigma_{i})\varrho_i(\sigma_{i},\sigma_{i+1})\nonumber \\
 & \times w(\sigma_{i+1},\sigma_{i+2})\ldots w(\sigma_{N},\sigma_{1}),
\label{eq:rho-df}
\end{align}
which can be alternatively rewritten as
\begin{equation}
\rho_{i}=\frac{1}{Z_{N}}\mathrm{Tr}\left(W^{i-1}PW^{N-i}\right)=\frac{1}{Z_{N}} \mathrm{Tr}\left(P W^{N-1}\right),
\end{equation}
where
\begin{equation}
P=\left(\begin{array}{cc}
\varrho_{i}(\tfrac{1}{2},\tfrac{1}{2}) & \varrho_{i}(\tfrac{1}{2},-\tfrac{1}{2})\\
\varrho_{i}(-\tfrac{1}{2},\tfrac{1}{2}) & \varrho_{i}(-\tfrac{1}{2},-\tfrac{1}{2})
\end{array}\right).
\end{equation}
In the thermodynamic limit ($N\rightarrow\infty$) the individual matrix elements of the partially averaged reduced density operator can be obtained after straightforward albeit cumbersome algebraic
calculation
\begin{align}
\rho_{i}= & \frac{1}{\Lambda_{+}}\left\{ \tfrac{\varrho_{i}(\tfrac{1}{2},\tfrac{1}{2})+\varrho_{i}(-\tfrac{1}{2},-\tfrac{1}{2})}{2}+\tfrac{2\varrho_{i}(\tfrac{1}{2},-\tfrac{1}{2})w_{+-}}{Q}\right.\nonumber \\
 & \left.+\tfrac{\left(\varrho_{i}(\tfrac{1}{2},\tfrac{1}{2})-\varrho_{i}(-\tfrac{1}{2},-\tfrac{1}{2})\right)\left(w_{++}-w_{--}\right)}{2Q}\right\} .\label{eq:rho-elem}
\end{align}
All the elements of the reduced density operator are consequently given by Eq. \eqref{eq:rho-elem}, so the thermally averaged reduced density operator can be expressed as follows
\begin{equation}
\rho_i=\left(\begin{array}{cccc}
\rho_{11} & 0 & 0 & \rho_{14}\\
0 & \rho_{22} & \rho_{23} & 0\\
0 & \rho_{32} & \rho_{33} & 0\\
\rho_{14} & 0 & 0 & \rho_{44}
\end{array}\right).\label{eq:rho-mat}
\end{equation}
It is worth noticing that the reduced density operator \eqref{eq:rho-mat} was already obtained in Ref. \cite{moises-pra}. Alternatively, one can use the free energy given by eq. \eqref{eq:free} to obtain the elements of the reduced density operator \eqref{eq:rho-mat} as a derivative with respect to the Hamiltonian parameters, such as discussed in Refs. \cite{strk-lyr}.

\section{Thermal entanglement}

Quantum entanglement is a peculiar type of correlation, which only emerges in quantum systems. The bipartite entanglement reflects nonlocal
correlations between a pair of particles, which persist even if they are far away from each other. To measure the bipartite entanglement
within the interstitial Heisenberg dimers in the spin-1/2 Ising-XYZ diamond chain we will exploit the quantity concurrence proposed by Wooters \textit{et al}. \cite{wootters,hill}.
The concurrence can be defined through the matrix $\mathbf{R}$
\begin{equation}
\mathbf{R}=\rho_i\left(\sigma^{y}\otimes\sigma^{y}\right)\rho_i^{*}\left(\sigma^{y}\otimes\sigma^{y}\right).\label{eq:R}
\end{equation}
which is constructed from the partially averaged reduced density operator given by Eq. (\ref{eq:rho-mat}) with $\rho_i^{*}$ being the complex conjugate of matrix $\rho_i$.
After that, the concurrence of the interstitial Heisenberg dimer can be obtained from the eigenvalues of a positive Hermitian matrix $\mathbf{R}$ defined by Eq.~\eqref{eq:R} from
\begin{equation}
\mathcal{C}=\mathrm{max}\{\sqrt{\lambda_{1}}-\sqrt{\lambda_{2}}-\sqrt{\lambda_{3}}-\sqrt{\lambda_{4}},0\},\label{eq:Cdf}
\end{equation}
where the individual eigenvalues are sorted in descending order $\lambda_{1}\geqslant\lambda_{2}\geqslant\lambda_{3}\geqslant\lambda_{4}$.
It can be shown that Eq.~(\ref{eq:Cdf}) can be reduced to the condition
\begin{equation}
\mathcal{C}=2\mathrm{max}\{|\rho_{23}|-\sqrt{\rho_{11}\rho_{44}},|\rho_{14}|-\sqrt{\rho_{22}\rho_{33}},0\},
\label{eq:conc-def}
\end{equation}
which depends on the matrix elements of the averaged reduced density operator given by Eq. \eqref{eq:rho-elem}.

\subsection{Concurrence}

The concurrence as a measure of the bipartite entanglement can be studied according to Eq. (\ref{eq:conc-def}) as a function of the coupling constants, external magnetic field and temperature.
\begin{figure}
\includegraphics[scale=0.31]{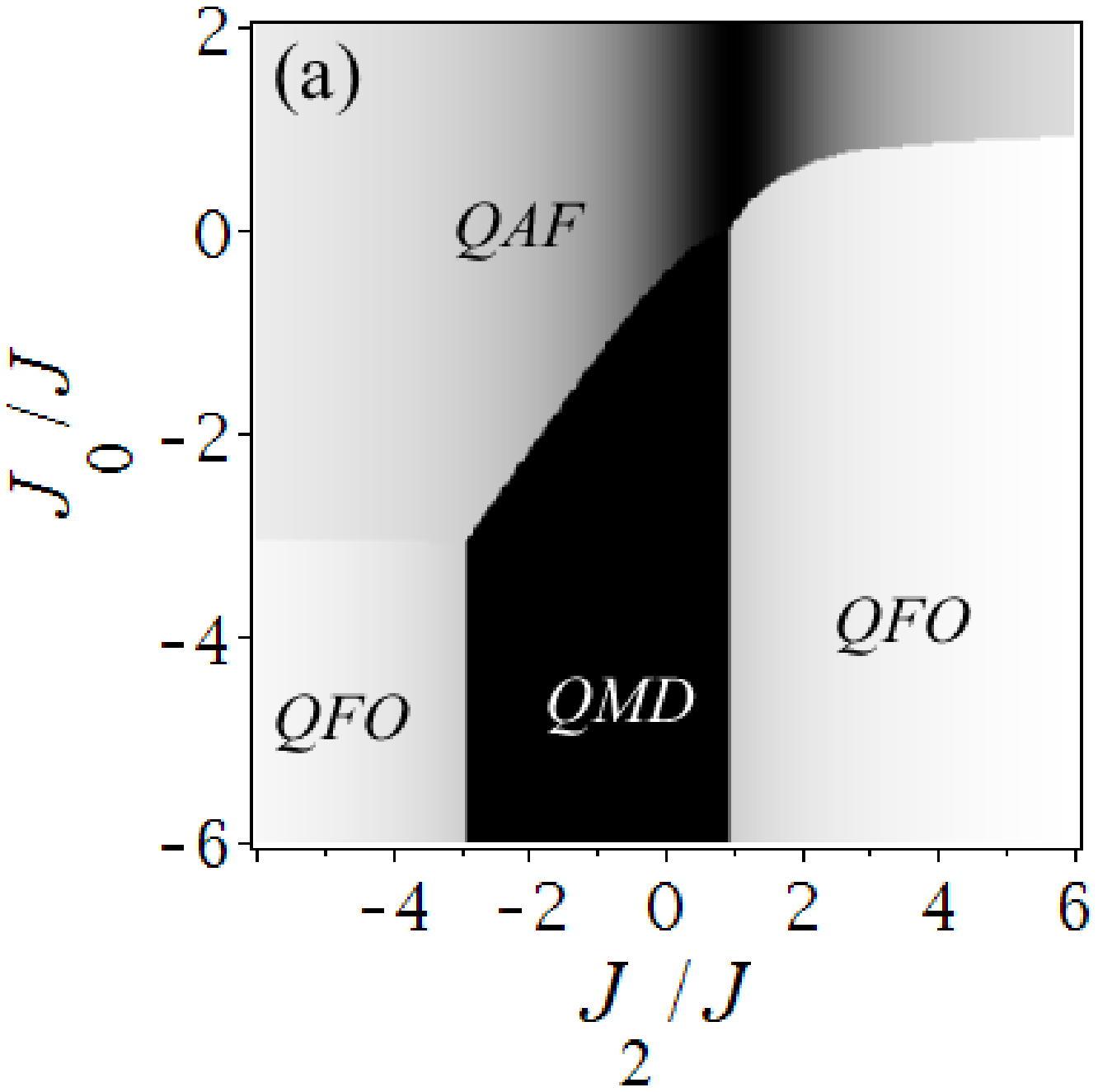}\vspace{-0.3cm}\includegraphics[scale=0.31]{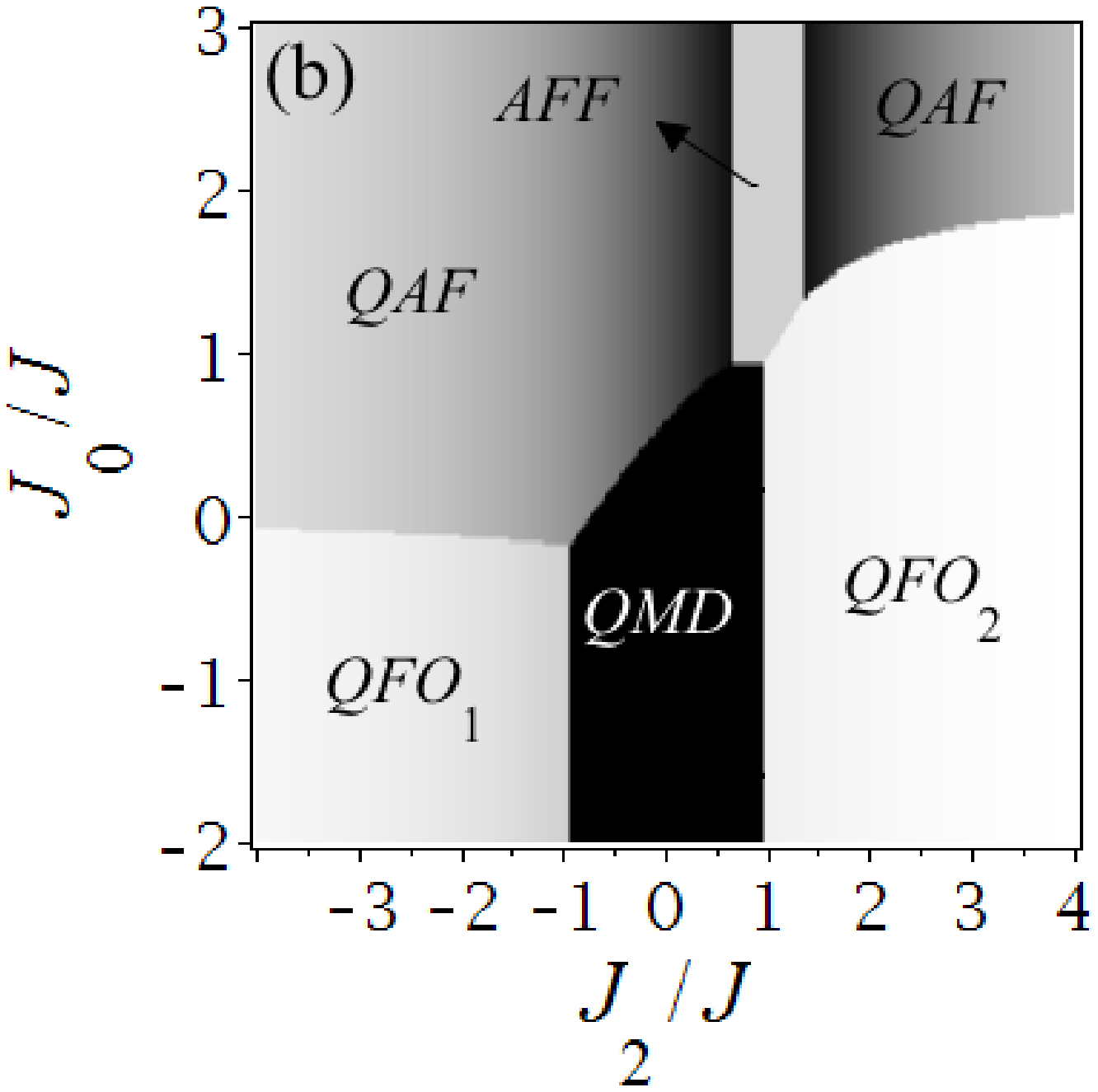}\caption{\label{fig:J0vsJ2}The density plot of concurrence for $J_{1}/J=J_{z}/J=1$, $\gamma=0.5$, $k_BT/J=0$ and two different magnetic fields: (a) $h/J=0$; (b) $h/J=1$. }
\end{figure}
In Fig.\ref{fig:J0vsJ2} we depict the density plot of the concurrence
$\mathcal{C}$ as a function of $J_{0}/J$ and $J_{2}/J$ for a fixed
value of $J_{1}/J=1$, $J_{z}/J=1$, $\gamma=0.5$ and $k_BT/J=0$.
Black color corresponds to a maximum entanglement of the interstitial Heisenberg dimers ($\mathcal{C}=1$),
while white color stands for a disentangled character of the interstitial Heisenberg dimers  ($\mathcal{C}=0$).
Gray color implies a partial bipartite entanglement of the interstitial Heisenberg dimers ($0<\mathcal{C}<1$).
In Fig.\ref{fig:J0vsJ2}a we display the concurrence $\mathcal{C}$
by assuming zero magnetic field $h/J=0$ (c.f. with the ground-state phase diagram shown in Fig.\ref{fig:J0J2}a).
In the $QAF$ phase one observes the maximal entanglement of the interstitial Heisenberg spins
just around $J_{2}/J=1$, while the concurrence $\mathcal{C}$ decreases within the $QAF$ phase for any other $J_{2}/J$. Moreover, one can see
that the concurrence indicates a full bipartite entanglement within the $QMD$ phase in contrast with a relatively weak partial entanglement of the $QFO$ phase.

The density plot of the concurrence $\mathcal{C}$ is illustrated in Fig.\ref{fig:J0vsJ2}b as a function of $J_{0}/J$ and $J_{2}/J$ by
considering a non-zero magnetic field $h/J=1$ (c.f. with the ground-state phase diagram shown in Fig.\ref{fig:J0J2}b). Here, we can observe
presence of the $AFF$ phase with a weak concurrence $\mathcal{C}\lesssim0.2$, which arises on account of non-zero external magnetic field. Similarly, a small but non-zero concurrence can be detected within the $QFO_{1}$ and $QFO_{2}$ phases, which display just a weak entanglement of the interstitial Heisenberg spins. By contrast, the concurrence $\mathcal{C}\approx 1$ implies a very strong entanglement between the interstitial Heisenberg spins within the $QMD$ phase.

\begin{figure}
\includegraphics[scale=0.32]{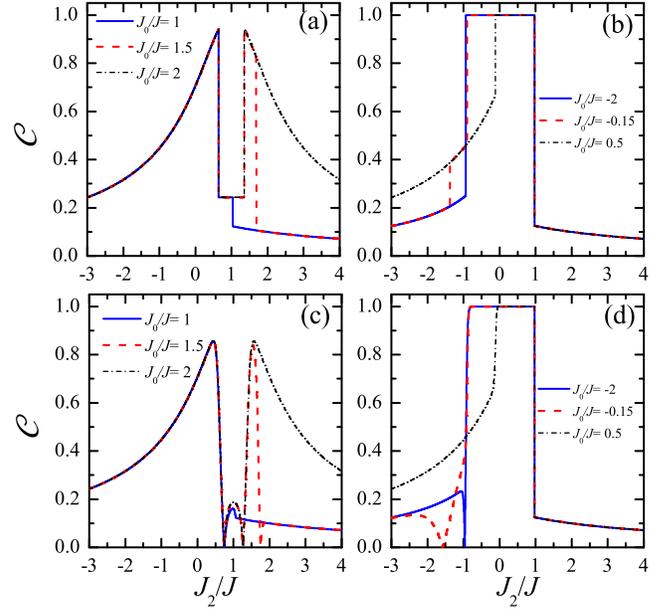}\vspace{-0.3cm}
\caption{\label{fig:CvsJ2} (Color online) (a)-(b) The concurrence $\mathcal{C}$ as a function of the ratio $J_{2}/J$ for $J_{1}/J=1$, $J_{z}/J=1$ , $h/J=1$,
$\gamma=0.5$, $k_BT/J=0$ and several values of $J_0/J$: (a) $J_0/J=1.0$, $1.5$ and $2.0$; (b) $J_0/J=0.5$, $-0.15$ and $-2.0$. (c)-(d) The plots of the concurrence
for the same set of the parameters as presented in figures (a)-(b) by assuming small but non-zero temperature $k_BT/J=0.01$.}
\end{figure}

Furthermore, the concurrence is depicted in Fig.\ref{fig:CvsJ2} against $J_{2}/J$ at zero and sufficiently low temperature
for the fixed values of $J_{1}/J=J_{z}/J=1$, $h/J=1$, $\gamma=0.5$ and a few different values of $J_{0}/J$. In the Fig.\ref{fig:CvsJ2}a,
the curve for $J_{0}/J=2$ (dot-dashed line) displays two different maxima of the concurrence
$\mathcal{C}\approx0.9417$ at $J_{2}/J=0.6438$ and $J_{2}/J=1.3562$, whereas the concurrence
approaches much lower value $\mathcal{C}\approx0.2425$ in the intermediate range $0.6438\leqslant J_{2}/J\leqslant1.3562$
corresponding to the $AFF$ phase. On the other hand, the concurrence decreases within the $QAF$ phase in the parameter region $J_{2}/J<0.6438$ and $J_{2}/J>1.3562$
as $J_{2}/J$ moves away from the ground-state boundaries with the $AFF$ phase.
For $J_{0}/J=1.5$ (dashed line) the concurrence behaves quite similarly to the aforedescribed
dot-dashed curve for $J_{0}/J=2$ up to $J_{2}/J\approx1.667$ because of presence of the same phases. However,
the concurrence then suddenly drops down to $\mathcal{C}\lesssim0.106$ for $J_{2}/J\gtrsim1.667$ due a zero-temperature phase transition from the highly entangled $QAF$ phase to the weakly entangled $QFO_{2}$ phase. The behavior of the concurrence for $J_{0}/J=1$ (solid line) is very similar to the previous cases $J_{0}/J=2$ (dot-dashed line) and $J_{0}/J=1.5$ (dashed line)
until $J_{2}/J=1$ is reached. The concurrence then shows a sudden drop at $J_{2}/J=1$ due to a zero-temperature phase transition from the $AFF$ phase to the $QFO_{2}$ phase,
where the concurrence monotonically decreases from $\mathcal{C}\approx0.123$ as the ratio $J_{2}/J$ further strengthens.

In Fig.\ref{fig:CvsJ2}b the concurrence at first steadily increases for $J_{0}/J=-2$ (solid line) within the parameter region corresponding to the $QFO_{1}$ phase. The quantum phase transition between the $QFO_{1}$ and $QMD$ phases is accompanied with an abrupt stepwise change of the concurrence from $\mathcal{C}\approx0.25$ up to its highest possible value $\mathcal{C}=1$, which is kept within the parameter region belonging to the $QMD$ phase $-0.944 \lesssim J_{2}/J \lesssim 0.956$. The concurrence then suddenly decreases at the ground-state boundary between the $QMD$ and $QFO_{2}$ phases at $J_{2}/J\approx0.956$ down to $\mathcal{C} \approx 0.125$ and from this point it shows a gradual monotonous decline upon further increase of the ratio $J_{2}/J$. Analogously, another particular case for $J_{0}/J=-0.15$ (dashed line) also exhibits a gradual increase of the concurrence within the phase $QFO_{1}$ until $\mathcal{C}\approx0.2034$ is reached at $J_{2}/J\approx-1.378$. However, the concurrence then suddenly increases to $\mathcal{C}\approx0.388$ at $J_{2}/J\approx-0.944$ due to a quantum phase transition from the $QFO_{1}$ phase to the $QAF$ phase, which persists at moderate values of $J_2/J$. The concurrence then suddenly jumps to its maximum value $\mathcal{C}=1$ at $J_{2}/J\approx-0.910$ owing to another quantum phase transition between the $QAF$ and $QMD$ phases. At higher values of the interaction ratio $J_{2}/J\gtrsim-0.910$ the concurrence behaves for $J_{0}/J=-0.15$ quite similarly to the aforedescribed case $J_{0}/J=-2$ (cf. solid and dashed line). The most notable difference for the last depicted particular case $J_{0}/J=0.5$ (dash-dot line) lies in absence of the $QFO_{1}$ phase. As a result, the concurrence for $J_{0}/J=0.5$ steadily increases within the $QAF$ phase to $\mathcal{C}\approx2/3$ achieved at $J_{2}/J\approx-0.127$, where the concurrence abruptly jumps to its maximum value $\mathcal{C}=1$ because of a quantum phase transition from the $QAF$ phase to the $QMD$ phase. The maximum value of the concurrence is retained until the ground-state phase boundary between the $QMD$ and $QFO_{2}$ phases is reached, where the concurrence suddenly drops to $\mathcal{C}\approx1/8$
at $J_{2}/J\approx0.956$. Consequently, the concurrence shows the same trends for $J_{2}/J\gtrsim-0.127$ for all three aforementioned cases.

To gain an insight into the effect of temperature, the concurrence $\mathcal{C}$ is displayed in Fig.\ref{fig:CvsJ2}c as a function of $J_{2}/J$ by assuming small but non-zero temperature $k_BT/J=0.01$ for the same set of parameters as used in Fig.\ref{fig:CvsJ2}a. It is quite obvious from a comparison of Figs. \ref{fig:CvsJ2}a and c that the concurrence shows a pronounced changes upon small variation of temperature just within the $AFF$ phase, while it is almost unaffected by small temperature fluctuations in all the other phases. The similar plot illuminating small temperature effect upon the concurrence is illustrated in Fig.\ref{fig:CvsJ2}d for the same set of parameters as used in Fig.\ref{fig:CvsJ2}b. As one can see, the concurrence is quite sensitive to small temperature changes just if the interaction ratio $J_{0}/J$ is selected sufficiently close to the ground-state phase boundary between the $QFO_1$ and $QAF$ phases, whereas it is quite robust with respect to small temperature fluctuations in the rest of parameter space.

\begin{figure}
 \includegraphics[scale=0.32]{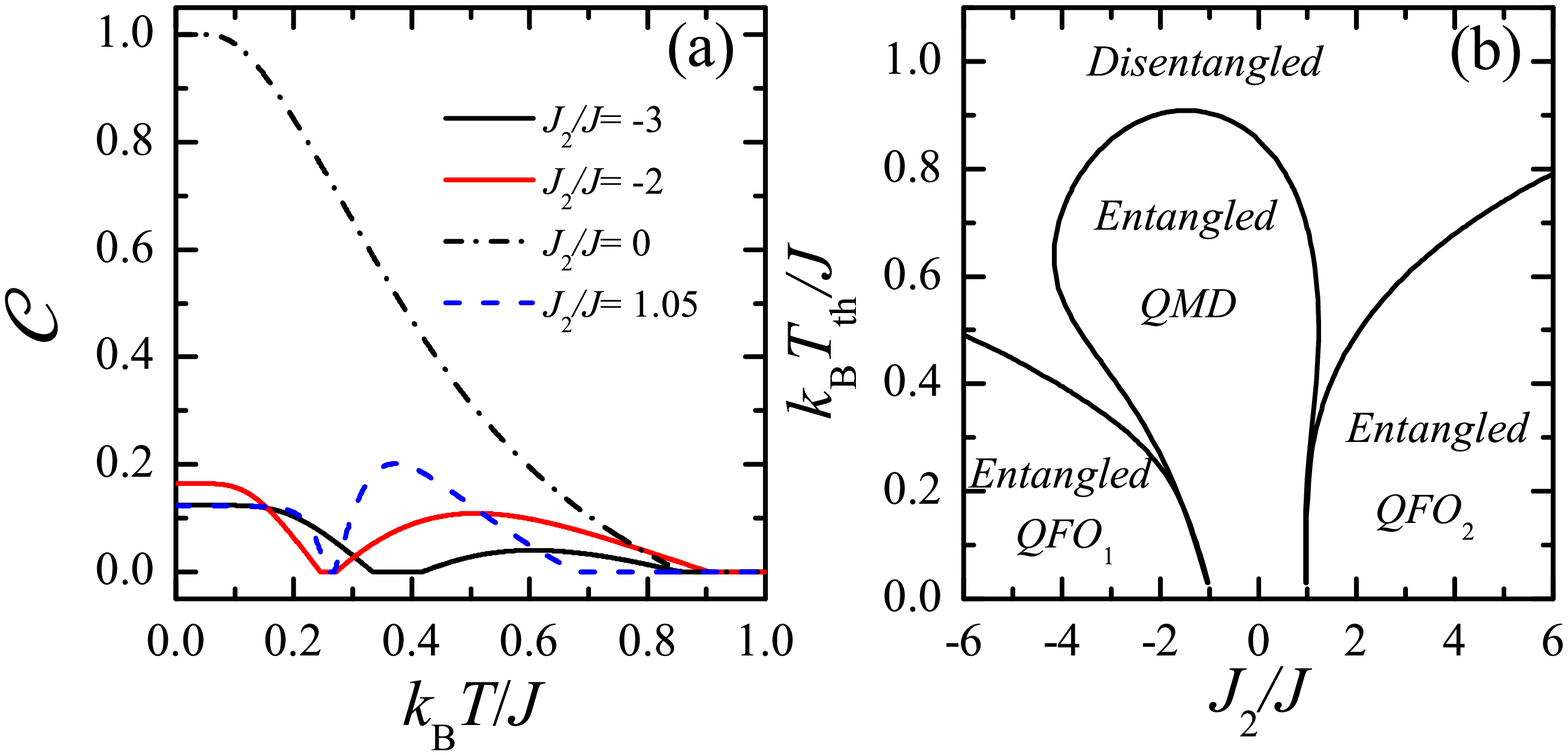}\vspace{-0.3cm}\caption{\label{fig:TvsCu} (Color online) (a) Temperature dependences of the concurrence for the fixed values of the coupling constants $J_{1}/J=1$, $J_{z}/J=1$, $J_{0}/J=-1$, $h/J=1$, $\gamma=0.5$ and several values of $J_2/J$; (b) The threshold temperature $k_B T_{th}/J$ as a function of the interaction ratio $J_2/J$ for the same set of the coupling constants as used in (a).}
\end{figure}

A few typical thermal variations of the concurrence are plotted in Fig.\ref{fig:TvsCu}a for the fixed values of the coupling constants $J_{1}/J=1$, $J_{z}/J=1$, $J_{0}/J=-1$, $h/J=1$, $\gamma=0.5$ and several values of the interaction ratio $J_{2}/J$. It can be seen from this figure that the concurrence for $J_{2}/J=0$ monotonically decreases from its maximum value $\mathcal{C}=1$ upon rising temperature until it completely vanishes at the threshold temperature $T_{th}/J\approx0.857$. This standard thermal dependence of the concurrence appears whenever the selected coupling constants drive the investigated system towards the $QMD$ ground state with a full entanglement of the interstitial Heisenberg dimers. On the other hand, Fig.\ref{fig:TvsCu}a also illustrates several more striking temperature dependences of the concurrence serving as a measure of the bipartite entanglement. For instance, the concurrence for $J_{2}/J=1.05$ diminishes from much lower initial value $\mathcal{C} \approx 0.122$ upon increasing temperature in agreement with a weaker entanglement of the $QFO_2$ phase. The concurrence disappears at first threshold temperature $k_B T_{1} /J \approx0.25$, then it re-appears at slightly higher second threshold temperature until it definitely vanishes at third threshold temperature
$k_B T_{3}/J\approx0.68$. The reentrance of the bipartite entanglement at higher temperatures can be attributed to a thermal activation of the $QMD$ phase. A similar thermal reentrance of the concurrence can be also detected for $J_{2}/J=-2.0$ and $-3.0$, but the entangled region at low temperatures now corresponds to the $QFO_1$ phase rather than to the $QFO_2$ phase.

To get an overall insight we have displayed in Fig.\ref{fig:TvsCu}b the threshold temperature $k_B T_{th}/J$ as a function of the interaction ratio $J_{2}/J$ for the same set of the coupling constants $J_{1}/J=1$, $J_{z}/J=1$, $J_{0}/J=-1$, $h/J=1$ and $\gamma=0.5$ as discussed before. It is noteworthy that the left (right) wing of the threshold temperature delimits the bipartite entanglement above the $QFO_1$ ($QFO_2$) ground state, while the intermediate loop region delimits the bipartite entanglement above the $QMD$ ground state. The intermediate closed loop starts at zero temperature from asymptotic limits $J_{2}/J\approx-0.944$ and $0.956$, but this region generally spreads over a wider range of the parameter space at higher temperatures on account of a stronger entanglement of the $QMD$ phase in comparison with that of the $QFO_1$ and $QFO_2$ phases. Owing to this fact, the reentrance of the concurrence emerges in a close vicinity of both quantum phase transitions $QMD-QFO_1$ and $QMD-QFO_2$. Note that reentrant transitions between the entangled and disentangled states have been reported on previously also for another quantum spin systems \cite{Morri}. For completeness, let us mention that the entanglement due to the $QFO_{1}$ and $QFO_{2}$ phases originates from the expression $|\rho_{14}|-\sqrt{\rho_{22}\rho_{33}}$
of Eq.(\ref{eq:conc-def}), whereas the entanglement due to the $QMD$ phase comes from the other expression $|\rho_{2,3}|-\sqrt{\rho_{1,1}\rho_{4,4}}|$ of Eq.(\ref{eq:conc-def}).

\begin{figure}
\includegraphics[scale=0.32]{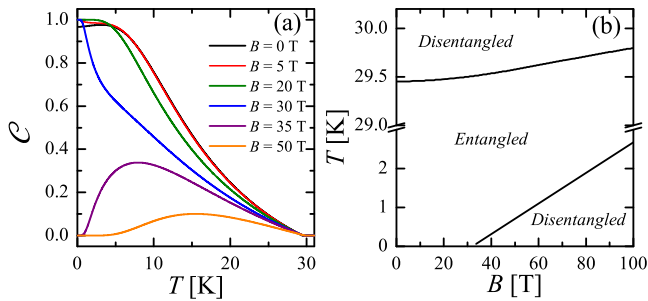}\vspace{-0.3cm}
\caption{\label{fig:CvsT}(Color online) (a) The temperature dependence of the concurrence for several values of the magnetic field $B$ and the coupling constants $J/k_B=J_{z}/k_B=33$K, $J_{1}/k_B=15.5$K, $J_{2}/k_B=6.9$K, $J_{0}/k_B=4.6$K, $\gamma=0$, $g=2.06$ relevant to the azurite; (b) The threshold temperature as a function of magnetic field $B$ for the coupling constants relevant to the azurite.}
\end{figure}

In the following we will study the concurrence $\mathcal{C}$ of the spin-1/2 Ising-Heisenberg diamond chain adapted by fixing the coupling constants $J/k_B=J_{z}/k_B=33$~K, $J_{1}/k_B=15.5$~K, $J_{2}/k_B=6.9$~K, $J_{0}/k_B=4.6$~K, $\gamma=0$ and $g=2.06$ in order to describe the bipartite entanglement of the natural mineral azurite Cu$_{3}$(CO$_{3}$)$_{2}$(OH)$_{2}$. Thermal variations of the concurrence $\mathcal{C}$ are plotted in Fig.\ref{fig:CvsT}a for several values of the magnetic field $B$. If the magnetic field is selected below the saturation value $B<B_s \approx 32$~T, then, the concurrence monotonically  decreases with increasing temperature until it completely vanishes at the threshold temperature $T_{th} \approx 30$~K. Moreover, it can be observed from Fig.\ref{fig:CvsT}a that the closer the magnetic field to its saturation field is, the steeper the decline of the concurrence is. On the other hand, the thermal behavior of the concurrence fundamentally changes when the magnetic field exceeds the saturation value $B>B_s$. Under this condition, the concurrence is at first zero at low enough temperatures due to a classical character of the SPA phase, then it becomes non-zero at a lower threshold temperature and finally disappears at the upper threshold temperature $T_{th} \approx 30$~K. It is noteworthy that the lower threshold temperature monotonically increases as the magnetic field is lifted from its saturation value (see the bottom part of Fig.\ref{fig:CvsT}b). On the other hand, it is quite surprising that the upper threshold temperature exhibits only a weak dependence on the magnetic field within the range of accessible magnetic fields (see the upper part of Fig.\ref{fig:CvsT}b). It could be thus concluded that the azurite Cu$_{3}$(CO$_{3}$)$_{2}$(OH)$_{2}$ remains thermally entangled below the threshold temperature $T_{th} \approx 30$~K regardless of the applied magnetic field.

\section{Conclusions}

The present article is devoted to the spin-1/2 Ising-XYZ distorted diamond chain accounting for the nearest-neighbor XYZ coupling between the interstitial spins, the nearest-neighbor Ising coupling between the nodal and interstitial spins, as well as, the second-neighbor Ising interaction between the nodal spins. Magnetic and thermodynamic properties of the model under investigation have been exactly calculated using the transfer-matrix technique. Besides the quantum antiferromagnetic phase QAF and the quantum monomer-dimer phase QMD, the ground-state phase diagram of the spin-1/2 Ising-XYZ distorted diamond chain may involve due to the XY anisotropy two unprecedented quantum ferromagnetic phases $QFO_{1}$ and $QFO_{2}$, as well as, the modulated antiferromagnetic-ferromagnetic phase $AFF$. In particular, our attention has been paid to a rigorous analysis of the quantum entanglement of the interstitial Heisenberg dimers at zero as well as non-zero temperatures through the concurrence serving as a measure of the bipartite entanglement. It has been demonstrated that the concurrence may exhibit either standard thermal dependence with a monotonous decline with increasing temperature or a more peculiar non-monotonous thermal dependence with a reentrant rise and fall of the concurrence (thermal entanglement). In addition, it is conjectured that the bipartite entanglement between the interstitial Heisenberg dimers in the natural mineral azurite  Cu$_{3}$(CO$_{3}$)$_{2}$(OH)$_{2}$ is quite insensitive to the applied magnetic field and it persists up to approximately 30 Kelvins.

\section*{Acknowledgment}

O. Rojas, M. Rojas and S. M. de Souza thank CNPq and FAPEMIG for partial financial support. J. Torrico thank CAPES for partial financial support. M. L. Lyra thank CAPES, CNPq and FAPEAL for partial financial support. J. Stre\v{c}ka acknowledges the financial support provided under the grant No. VEGA 1/0043/16.


\begin{thebibliography}{50}
\bibitem{AmicoHorod}L. Amico, R. Fazio, A. Osterloh and V. Vedral,
Rev. Mod. Phys. \textbf{80,} 517 (2008); R. Horodecki et al., Rev.
Mod. Phys. \textbf{81,} 865 (2009); O. Gühne and G. Tóth, Phys. Rep.
\textbf{474}, 1 (2009).

\bibitem{nielsen}M. A. Nielsen and I. L. Chuang, \textit{Quantum
Computation and Quantum Information}, Cambridge University Press, 2010.

\bibitem{qubit-Heisnb}K. M. O'Connor and W. K. Wootters, Phys. Rev.
A \textbf{63}, 052302 (2001); Y. Sun, Y. Chen and H. Chen, Phys. Rev.
A \textbf{68}, 044301 (2003); G. L. Kamta and A. F. Starace, Phys.
Rev. Lett. \textbf{88}, 107901 (2002).

\bibitem{kikuchi05}H. Kikuchi, Y. Fujii, M. Chiba, S. Mitsudo, T.
Idehara, T. Tonegawa, K. Okamoto, T. Sakai, T. Kuwai, and H. Ohta,
Phys. Rev. Lett. \textbf{94}, 227201 (2005).

\bibitem{Lau} A. Honecker and A. Lauchli. Phys. Rev. B \textbf{63,}
174407 (2001); H. Jeschke, I. Opahle, H. Kandpal, R. Valentí, H. Das,
T. Saha-Dasgupta, O. Janson, H. Rosner, A. Brühl, B. Wolf, M. Lang,
J. Richter, Sh. Hu, X. Wang, R. Peters, T. Pruschke, and A. Honecker,
Phys. Rev. Lett. \textbf{106}, 217201 (2011);
N. Ananikian, H. Lazaryan, and M. Nalbandyan, Eur. Phys. J. B \textbf{85,} 223 (2012).

\bibitem{honecker}A. Honecker, S. Hu, R. Peters J. Richter, J. Phys.: Condens. Matter \textbf{23,} 164211 (2011).

\bibitem{Cano}L. \v{C}anov\'a, J. Stre\v{c}ka, and M. Ja\v{s}\v{c}ur, J. Phys: Condens.
Matter \textbf{18}, 4967 (2006).

\bibitem{valverde}J. S. Valverde, O. Rojas, S.M. de Souza, J. Phys.:
Condens. Matter\textbf{ 20,} 345208 (2008); O. Rojas, S.M. de Souza,
Phys. Lett. A \textbf{375,} 1295 (2011).

\bibitem{orojas}O. Rojas, S. M. de Souza, V. Ohanyan, M. Khurshudyan,
Phys. Rev. B \textbf{83} , 094430 (2011).

\bibitem{lis-1}B. M. Lisnii, Ukr. J. Phys. \textbf{56}, 1237 (2011).

\bibitem{mrojas}O. Rojas, M. Rojas, N. S. Ananikian and S. M. de
Souza, Phys. Rev. A \textbf{86}, 042330 (2012).

\bibitem{mrojas-1}J. Torrico, M. Rojas, S. M. de Souza, Onofre Rojas
and N. S. Ananikian, Europhysics Letters \textbf{108}, 50007 (2014).

\bibitem{mrojas2} J. Torrico, M. Rojas, M. S. S. Pereira, J. Strecka
and M. L. Lyra, Phys. Rev. B \textbf{93}, 014428 (2016).

\bibitem{moises-aphy} M. Rojas, S. M. de Souza and O. Rojas, Ann.
Phys., \textbf{377}, 506 (2017).

\bibitem{cheng} W. W. Cheng, X.Y. Wang, Y.B. Sheng, L.Y. Gong, S.M. Yhao, J.M. Liu, Sci. Rep. \textbf{7}, 42360 (2017).

\bibitem{stre} B. Lisnyi and J. Stre\v{c}ka, Phys. Status Solidi B \textbf{251}, 1083 (2014).

\bibitem{Faizi}E. Faizi , H. Eftekhari , Rep. Math. Phys., 74, 251
(2014).

\bibitem{vahan}V. Hovhannisyan, J. Stre\v{c}ka, N. Ananikian, J. Phys.:
Condens. Matter \textbf{28}, 085401 (2016).

\bibitem{Fisher} M. Fisher, Phys. Rev. \textbf{113,} 969 (1959).

\bibitem{Syozi}I. Syozi, Prog. Theor. Phys.\textbf{ 6,} 341 (1951).

\bibitem{phys-A-09}O. Rojas, J. S. Valverde, S. M. de Souza, Physica
A \textbf{388,} 1419 (2009); O. Rojas, S. M. de Souza, J. Phys. A:
Math. Theor. \textbf{44,} 245001 (2011).

\bibitem{moises-pra}O. Rojas, M. Rojas, N. S. Ananikian, S. M. de
Souza, Phys. Rev. A \textbf{86}, 042330 (2012).

\bibitem{strk-lyr}J. Stre\v{c}ka, O. Rojas, T. Verkholyak, M. L. Lyra,
Phys. Rev. E \textbf{89}, 022143 (2014)

\bibitem{wootters}W. K. Wootters, Phys. Rev. Lett. \textbf{80}, 2245
(1998).

\bibitem{hill}S. Hill and W.K. Wootters, Phys. Rev. Lett. \textbf{78},
5022 (1997).

\bibitem{Morri}S. Morrison and A. S. Parkins, Phys. Rev. Lett. \textbf{100},
040403 (2008).
\end{thebibliography}
\end{document}